# Hypersampling of pseudo-periodic signals by analytic phase projection

Henning U. Voss

Department of Radiology

Weill Cornell Medicine

Citigroup Biomedical Imaging Center, 516 E 72nd Street

New York, NY 10021

Tel. 001-212-746-5216

Fax. 001-212-746-6681

Email: hev2006@med.cornell.edu

April 27, 2018

**Abstract**

A method to upsample insufficiently sampled experimental time series of pseudo-periodic signals is proposed. The result is an estimate of the pseudo-periodic cycle underlying the signal. This "hypersampling" requires a sufficiently sampled reference signal that defines the pseudo-periodic dynamics. The time series and reference signal are combined by projecting the time series values to the analytic phase of the reference signal. The resulting estimate of the pseudo-periodic cycle has a considerably higher effective sampling rate than the time series. The procedure is applied to time series of MRI images of the human brain. As a result, the effective sampling rate could be increased by three orders of magnitude. This allows for capturing the waveforms of the very fast cerebral pulse waves traversing the brain. Hypersampling is numerically compared to the more commonly used retrospective gating. An outlook regarding EEG and optical recordings of brain activity as the reference signal is provided.

## I. INTRODUCTION

Many signals in the biological and biomedical sciences are of a pseudo-periodic nature with irregularly spaced, stretched, or otherwise distorted variations of a repeating cycle. An example for a pseudo-periodic cycle is the characteristic QRS complex observed in blood pressure signals and electric recordings of the heart [1]. Another example are patterns of electrical activity of the brain observed in electroencephalographic (EEG) surface recordings [2]. Those signals usually can be measured with a sufficient sampling rate to resolve their underlying pseudo-periodic cycles (QRS-complex, EEG waveform, respectively). However, often it is not possible to measure the effects of the pseudo-periodic dynamics in parts of the body that cannot be accessed so easily, for example deep within the brain. The method of choice to obtain signals from anywhere in the brain is MRI. Dynamic or functional MRI of the brain is typically sampled at an insufficient rate to resolve the cardiac cycle or EEG patterns [3]. In order to investigate the pseudo-periodic signal in a particular location within the brain, one solution is to upsample the MRI signal at that point with an effective sampling time that is much smaller than the average cardiac cycle or EEG waveform period. The cardiac or EEG recordings then can serve as a reference used to define the pseudo-periodicity of the dynamics of interest.

Here, an efficient upsampling procedure, called hypersampling, is described. Hypersampling consists of an upsampling of the undersampled time series by using the method of analytic phase projection (APP). Hypersampling can be seen as a generalization of retrospective gating [4, 5]. Whereas in retrospective gating a recurring template is identified from the reference signal, in hypersampling the continuous phase underlying the pseudo-periodic reference signal is identified from the reference signal. This phase estimate is then used for APP.

The organization of this manuscript is as follows: First, hypersampling by APP is described in Section II. Hypersampling is demonstrated on simulations in Section III. In Section IV, these concepts are applied to magnetic resonance imaging (MRI) of the brain in order to visualize the very fast pulse waves traversing the brain, which normally cannot be resolved with MRI. A discussion including a comparison with retrospective gating and possible further applications to other hybrid systems with fast and slow time scales concludes the manuscript. An appendix provides code for hypersampling via APP, and the supplementary data a video of the pulse waves observed in the human brain.

## II. METHOD

### A. Overview

Hypersampling by analytic phase projection (APP) is summarized in Fig. 1.

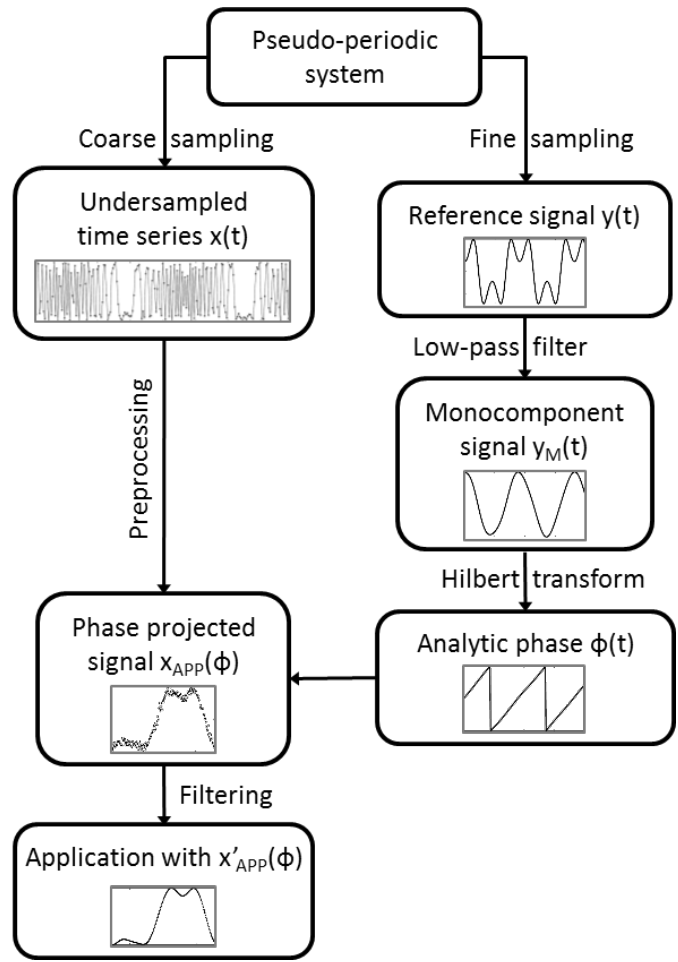

**Figure 1.** Schematics of the analytic phase projection (APP) method to upsample an undersampled signal. Please refer to text for details.

Two signals are acquired from the system under study: An undersampled pseudo-periodic time series $x(t)$ and a sufficiently sampled pseudo-periodic reference signal $y(t)$. The reference signal and the time series are acquired during the same time interval and are assumed to have the same pseudo-periodicity. Then, a monocomponent signal $y_M(t)$ is obtained by low-pass filtering the reference signal $y(t)$ (box "Monocomponent signal $y_M(t)$"). Monocomponent signals have a monotonically increasing phase. In other words, in a monocomponent signal the instantaneous frequency or time derivative of its phase is non-negative at any time [6]. This phase monotonicity is needed later on in the phase projection step, which requires the phase to be a piecewise invertible function. In practice, the phase of a signal can only be obtained modulo an interval of length $2\pi$, which causes phase resets. At phase resets, the phase has a discontinuity from a value near $\pi$ to a value near $-\pi$, thereby crossing the zero line. The phase is estimated from the monocomponent signal as its analytic phase (box "Analytic phase $\Phi(t)$"; two phase resets are visible). The analytic phase is interpolated to the time series sampling times. Then, the time series values are assigned to their corresponding phase values (box "Phase projected signal $x_{APP}(\Phi)$"). In other words, a coordinate transformation from time to phase is performed: Whereas the original time series depends on time, the phase-projected time series depends on the pseudo-periodic cycle phase. The time series values themselves are not altered, they are just re-ordered, thus the description as a "projection". The method is referred to as "hypersampling" because the main requirement is that the underlying pseudo-periodic process is sampled over a time that spans as many pseudo-periods as possible. Finally, depending on the particular application, it might be necessary to further filter the result in order to obtain an estimate for the phase-projected cycle (box "Application with $x'_{APP}(\Phi)$").

### B. Computational details

A monocomponent signal can be written as the product of an instantaneous amplitude $\rho(t) \geq 0$ and an instantaneous phase factor $\cos(\Phi(t))$, or as an amplitude-phase modulation [6, 7]. Writing the signal $y_M(t)$ as an amplitude-phase modulation

$$y_M(t) = \rho(t) \cos \Phi(t), \tag{1}$$

its analytic extension is

$$y_A(t) = y_M(t) + iy_H(t) = \rho(t)e^{i\Phi(t)}, \tag{2}$$

with the Hilbert transform [6]

$$y_H(t) = \frac{1}{\pi}P\int_R \frac{y_M(\tau)}{t-\tau}d\tau$$
$$= \lim_{\varepsilon \to 0^+}\frac{1}{\pi}\int_{|t-\tau|>\varepsilon} \frac{y_M(\tau)}{t-\tau}d\tau. \tag{3}$$

The integral in this expression is a principal value integral. The analytic extension (2) expressed via the Hilbert transform (3) provides a unique expression for the amplitude-phase modulation (1), the canonical amplitude-phase modulation [6]. The Hilbert transform itself can be computed by standard signal processing software [8]. The analytic phase follows from the analytic signal as

$$\Phi(t) = \arg(y_A(t)) = \arg(y_M(t) + i\, y_H(t)). \tag{4}$$

The argument function here is the four-quadrant inverse tangent relation, sometimes denoted $atan2(y_H(t), y_M(t))$. Its principal values are restricted to the interval $(-\pi, \pi]$. In a monocomponent signal, the instantaneous frequency is always non-negative, i.e., $d\Phi(t)/dt \geq 0$, for all time points where it is defined. Thus, the analytic phase is monotonically increasing, and decreasing only during phase resets. The phase monotonicity can be checked, for example, visually by graphing the analytic phase. If necessary, the low-pass filter can be adjusted such as to improve monotonicity of the analytic phase. Depending on the application, it might also be necessary to preprocess the time series, for example to remove trends. Once an approximately monotonic phase of the reference signal has been obtained, one can proceed with the APP, which combines the time series and the analytic phase of the reference signal:

The (preprocessed) pseudo-periodic time series $x(t)$ is sampled at times $t_i$. The sampling times of the analytic phase $\Phi(t)$ are denoted by $\tau_j$. The analytic phase projection is a coordinate transformation of the time series sampling times to the analytic phase,

$$APP: x(t_i) \to x_{APP}(\Phi_i). \tag{5}$$

The index $i$ assumes values from 1 to N, the number of time series samples. Here, $\Phi_i = \Phi(t_i)$ is the analytic phase $\Phi(t)$ numerically interpolated to the signal sampling time $t_i$. This interpolation should be quite accurate in general, as the analytic phase of a monocomponent signal is an approximately smooth function, if sufficiently sampled, except at phase resetting points. For phases near phase reset, the interpolation can become inaccurate and it might be necessary to provide corrective measures, for example discarding outliers. If to each time series sampling time $t_i$ there is a corresponding reference signal sampling time $\tau_j = t_i$, the interpolation step can be omitted, i.e., the phases $\Phi(\tau_j)$ are taken directly as $\Phi_i = \Phi(\tau_j = t_i)$.

The upsampled, phase projected signal $x_{APP}(\Phi)$ is an estimate of the pseudo-periodic signal averaged over one period, or, in other words, an estimate of the pseudo-periodic cycle. In order to graph this cycle, the phases are ordered from their minimum value near $-\pi$ to their maximum value near $\pi$, and the signal values are ordered accordingly. Whereas the original time series values $x(t)$ depend on time $t$, the phase projected values $x_{APP}(\Phi)$ depend on phase $\Phi$. Finally, the phase-projected signal can be low-pass filtered in order to remove residual signal components that cannot be accounted for by pseudo-periodicity and thus analytic phase projection. The result is a smooth signal estimate $x'_{APP}(\Phi)$. The effective sampling interval of the upsampled signal follows from the number of time series samples, N, and the average pseudo-period of the reference signal, T, as

$$\Delta T_{eff} = T/N. \tag{6}$$

This shows that the more samples are measured, the larger the achieved effective sampling rate. Of note, the effective sampling rate does not depend on the original sampling rate of the time series.

## III. SIMULATION

A signal is being simulated as a nonlinearly transformed chirp signal with linearly increasing frequency from the beginning to the end of the signal time series in order to simulate pseudo-periodic data. Added to the data is 10% white noise. The Appendix contains a Matlab script defining the used signals and the analysis in detail. Figure 2, row A, contains the given signal. The sampling time of the signal is 2 s. (Artificial physical units are introduced here in order to facilitate comparison with the application to MRI data in the next section.) The signal is undersampled and therefore, the shape of a cycle cannot be recognized from this data. This shape is shown in row B (before addition of white noise) and not available to the investigation. The sampling time of the underlying true signal in row B is 0.01 s, and one can see in row B, left panel, that the cycle period at the beginning of the time series is 0.91 s. From row B, right panel, follows a cycle period of 0.45 s, due to the definition of the chirp signal that contains a doubling of frequency from the beginning to the end of the signal.

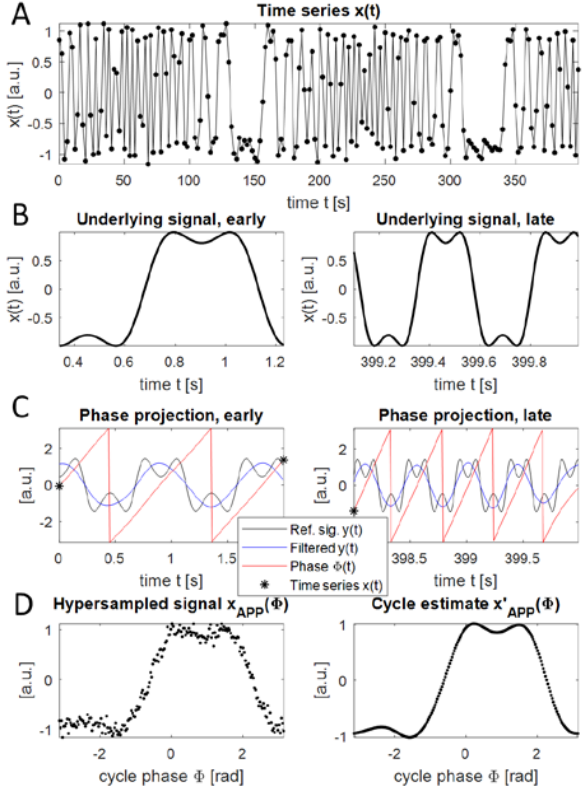

**Figure 2:** From the given undersampled signal containing pseudo-periodic data to the estimated hypersampled cycle. Please refer to text for a detailed description.

Row C shows the reference signal in black and the low-pass filtered reference signal in blue, again for the early and late signal portion. In the low-pass filtered signal, double peaks have been removed. This monocomponent signal enables a monotonic definition of the analytic phase, which is shown in red. The analytic phase is not unwrapped, i.e., only the principal values of the non-invertible arcus tangent relation in Eq. (4) are used. This causes phase resets at the end of each cycle. The first phase reset is seen at a time of 0.44 s in Fig. 2C, left panel. Importantly, the analytic phase is well defined both for low frequencies early on in the frequency-swept reference signal and for higher frequencies at the end of the reference signal (left and right panel in row C, respectively). In addition, the three sample times of the signal that happen to fall into the shown time intervals via a projection to the analytic phase are marked by an asterisk (*). This is the first part of the phase projection step; the signal sampling times are converted into phases of the cycle. In the final step, all phases are sorted and combined into one cycle. The result of the latter step, the hypersampled cycle, is shown in row D, left panel. The noise results mostly from the noise in the noise-contaminated original signal. Application of the same procedure to the underlying non-noisy signal would result in only little scatter, which is due to numerical inaccuracies (not shown). The right panel of row D shows a low-pass filtered hypersampled cycle. The latter one provides an accurate estimate of the original cycle (compare row B, left panel, with row D, right panel.

It is important to understand that the phase axis in Figure 2D does not have a time equivalent. The reason is that the signal consists of pseudo-periodic cycles of different lengths. However, one can define a "hypersampling rate" as the average sampling rate of the hypersampled cycle. In the present example, the signal contains 662 cycles with lengths from 0.45 to 0.91 s resulting from the frequency sweep from 2.21 to 1.10 Hz. With an average cycle length of 0.60 s and 200 data points in the original time series, the average sampling interval of the cycle would be about 3 ms. This is a reduction of the original signal sampling interval of 2 s by a factor of 662, the number of cycles contained in the time series.

## IV. APPLICATION: CEREBRAL PULSE WAVES

As an example, hypersampling via APP is applied to pulse waves in the human brain. Studying the properties of cerebral pressure waves and their possible effects on human health is a very active area of research [9-21]. Here, a functional MRI scan from a publicly available data base [22] is used to demonstrate that hypersampling can resolve pulse waveforms in the brain.

The MRI data consists of image volumes covering parts of frontal and occipital cortex and the regions in between, including an area with a dense distribution of main cerebral arteries (see Fig. 3B). The data is sampled every 2 s for 15 min at 1.2 mm isotropic voxel resolution. The image volumes are not further processed; processed data, such as motion corrected or template-brain-matched data, can cause artifacts that might be detrimental to this analysis [21]. Each voxel in the repeatedly sampled image volume contains a time series of 441 data points.

The reference signal consists of a pulse-oximetry signal acquired on a finger of the subject, sampled with 100 samples per second, or a sampling interval of 10 ms. This is a sufficient sampling rate to define the phase of the underlying pseudo-periodic process. The pseudo-period of the subject's cardiac dynamics, or the average RR interval, derived from the pulse signal, is 0.856 s. This is shorter than the MRI sampling interval of 2 s. Therefore, pulse waveforms contained in the MRI signal are not sampled sufficiently to resolve them.

First, the pulse reference signal is low-pass filtered with a cutoff frequency of 2 Hz in order to obtain an approximately monocomponent signal. The approximate monotonicity is validated by visual inspection of the filtered reference signal. Then, the exact acquisition times of the voxels under consideration are computed. This is important, as the whole data volume is acquired sequentially over a time interval of 2 s rather than at once. The time series are then high-pass filtered with a cutoff frequency of 0.0042 Hz in order to remove signal trends. The MRI signal is shown with negative sign to account for the fact that blood normally reduces the signal in this kind of MRI, if flow effects can be neglected.

The reference signal contains 1030 cycles. With an average pseudo-period of 0.856 s and 441 signal samples, the average sampling interval of the cardiac cycle follows from Eq. (6) as $\Delta T_{\text{eff}}$ = 1.9 ms. This is a reduction of the original signal sampling interval of 2 s by a factor of 1030, the number of cycles. Note that the original sampling rate of the MRI signal does not enter the calculation of $\Delta T_{\text{eff}}$.

Finally, the hypersampled time series is smoothed mildly with a cutoff frequency of 0.0083 Hz in order to remove signal scatter. This is the estimated pseudo-periodic cycle, or pulse waveform, in this particular application.

Figure 3A shows the waveform estimates for a point on the middle cerebral artery, as well as for the sagittal sinus, a vein. These two points are indicated in the pulse amplitude maps in Figs. 3B and C by numbers ① and ②, respectively. Those maps are defined in the caption of Fig. 3. The arterial waveform of Fig. 3A resembles aortic/carotid pulse waveforms, and thus has been annotated with terminology used in conventional pulse wave analysis [23-25]. The venous waveform is smaller (probably due to attenuation in the capillary bed) and less defined (probably due to dispersion) than the arterial waveform. An animation of the pulse wave amplitude (Fig. 3B) is available as supplemental information.

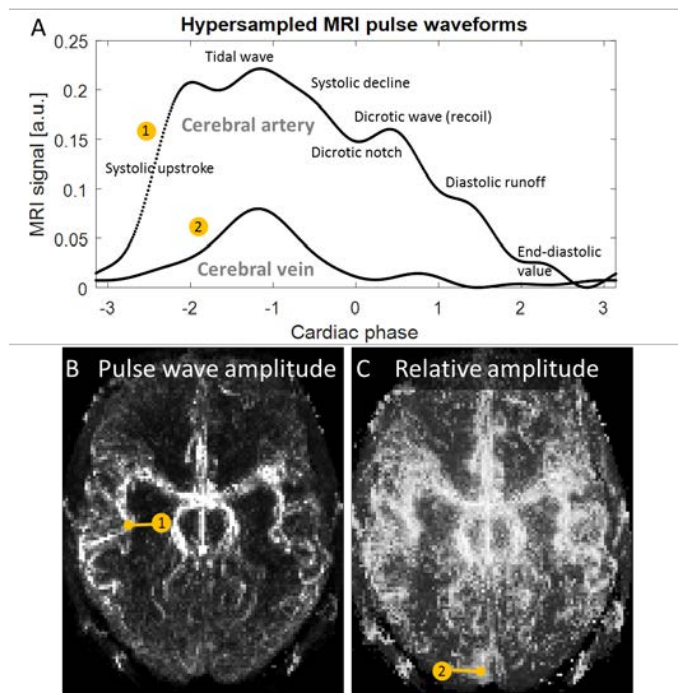

**Figure 3:** Hypersampled MRI signals of the brain resemble conventional pulse waveforms measured outside of the brain. Waveforms obtained from the middle cerebral artery and the sagittal sinus, a vein (A) and their location in maximum intensity projections of pulse wave amplitude (B, C). An animation of the pulse wave amplitude is available as supplemental information. Pulse wave amplitudes in panel B are defined as the maximum amplitude difference of the hypersampled waveforms and show mainly arteries. Pulse wave amplitudes in C, scaled by the standard deviation of the signal, also show areas with lower pulse wave amplitude, including venous components.

Finally, the average lag time between the arterial wave and the outgoing venous wave corresponds to the pulse wave transit time through the brain including the capillary bed. It is estimated as 105 ms, consistent with literature values of cerebral pulse transit times [26] derived from the brain periphery. If the average length of the vascular tree between those two points were known, the pulse wave velocity in the capillary bed could be estimated. Note that the pulse wave velocity is orders of magnitude higher than the blood flow velocity. For comparison, blood transit times would be several seconds [27].

## V. DISCUSSION

### A. *Comparison with retrospective gating*

The proposed method is generalizing well-known retrospective gating approaches. In most retrospective gating approaches, templates of the reference signal (for example, the R peak of the cardiac cycle) are used to define gating points. A pseudo-period, or cycle, is then defined by the interval from one template of the reference signal to the next, for example, RR intervals [4, 5]. Signal sampling times are then linearly projected on cycle times. The analytic phase projection method differs in two points.

First, it defines the pseudo-period without use of a template. This might be of particular importance for applications where template fitting could be erroneous, like in respiratory signals with a large variation in amplitude, and EEG signals with a large noise component. The present approach splits the reference signal dynamics into amplitude and phase dynamics, via the Hilbert transform, and then uses the phase evolution for timing. (Other pulse data analysis approaches involving the Hilbert transform [28-32] do not exploit the monocomponent properties of the signal but rather estimate analytic amplitudes of the signals or their derivatives. Those approaches still require some form of template matching.)

Second, the analytic phase takes account of local nonlinearities of the references signal phase evolution. The projection of time series sampling times to phases does not need to be linear but takes account of the possibly nonlinear cycle dynamics of the reference signal.

In the following, it is demonstrated on numerical simulations that hypersampling can outperform retrospective gating when the underlying signal phase varies nonlinearly. This is accomplished by first defining a cycle template and then distributing stretched and compressed copies of the cycle template over time with randomly varying intervals, or gaps, between cycles. Such a signal is shown in Figs. 4C and 5C. Application of hypersampling yields an estimate (Fig. 4D, right panel) that is comparable to the case without random gaps (Fig. 4B, right panel). For the signal of the latter, see Fig. 4A. Comparing these results to Fig. 5, retrospective gating [5], shows that in the case of gaps between the cycles the cycle estimate becomes inaccurate (Fig. 5D, right panel). It does not reflect the situation that there are gaps between the basic cycles, which actually belong to the cycle and make it a cycle with nonlinear phase variation. This phase variation becomes visible in the estimated analytic phase in Fig. 4C. This should be compared with Fig. 5C, which shows the linear gating intervals analogously to the phase.

One can also see in Fig. 4C that the filtering to obtain a monocomponent signal was not sufficient. There is a short interval with negative slope of the phase starting at around 8.5 s. This contributes to the scatter of the hypersampled signals (Figs. 4B, D, left panels). This scatter actually can be worse than for retrospective gating in case of phase linearity (Fig. 5B, left panel).

In conclusion, both methods have their potentials and limitations depending on the specific circumstances.

waveforms. The MRI signal units are denoted here as "arbitrary" as a clear relationship to blood volume, pressure, flow, and MRI scan parameters is still an ongoing research effort [17]. However, as any MRI signal is proportional to the imaged number of excited hydrogen spins, the blood volume is expected to be a dominant signal generating parameter. Blood pressure p and volume V in arteries are related by the arterial compliance $C(p) = dV(p)/dp$. For low pressure, the compliance is approximately constant, and thus the volume proportional to pressure. The compliance then levels off for larger pressure, depending on the arterial stiffness. Therefore, the observed waveforms could be an approximation for intracranial blood pressure within the linear regime. In addition, some analysis methods of arterial waveforms do not require absolute blood pressure values, for example methods that use the slopes of pressure over time [33] or the timing of peaks/troughs [25].

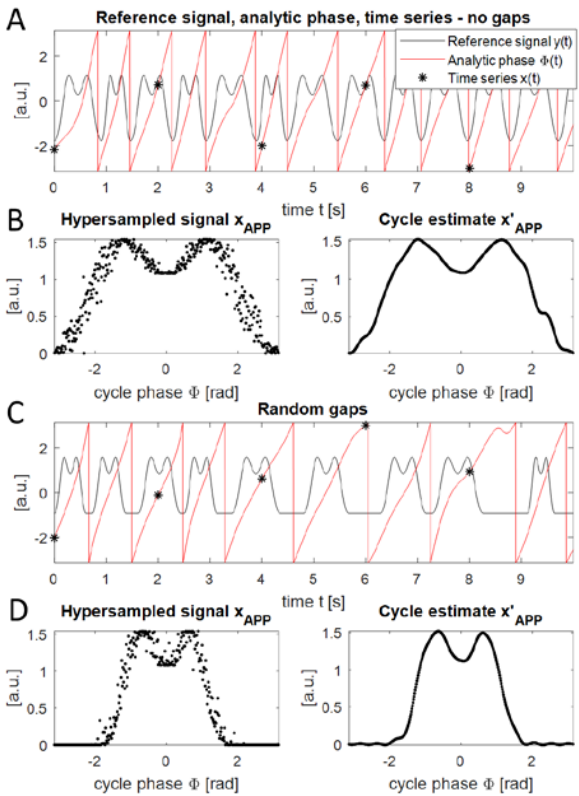

**Figure 4:** APP applied to simulated time series. To be compared with Fig. 5. A: The reference signal (here identical to the underlying, unknown signal to be estimated), the analytic phase, and time series samples, for a short section of the data. B: The hypersampled signal (left panel) and the cycle estimate (right panel). C: Same as A, but with randomly spaced cycles in the underlying signal. D: Same as B but for randomly spaced cycles. The cycle estimate still accurately reflects the underlying signal shape in panel C.

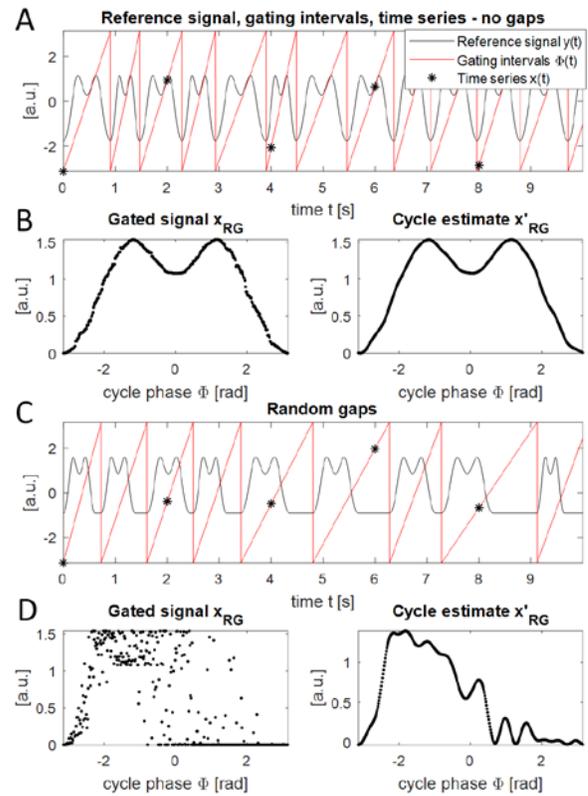

**Figure 5:** Retrospective gating applied to simulated time series. To be compared with Fig. 4. A: The reference signal (here identical to the underlying, unknown signal to be estimated), the gating intervals, and time series samples, for a short section of the data. B: The gated signal (left panel) and the cycle estimate (right panel). C: Same as A, but with randomly spaced cycles in the underlying signal. D: Same as B but for randomly spaced cycles. The cycle estimate does not accurately reflect the underlying signal shape in panel C anymore.

### B. Clinical research application

The results of Fig. 3, namely the clear identification of characteristic waveforms resembling aortic or carotid blood pressure waves, makes the MRI signal amenable to established procedures of waveform analysis, applied now to cerebral pulse

Those methods, which have proven significant with respect to the health of the cardiovascular system, could readily be applied to the hypersampled intracranial waveforms to investigate the status of the cerebrovascular system. The pulse transit time from the middle cerebral artery to the sagittal sinus estimated here already contains information of average pulse wave velocity and traveled distance. Whereas conventional (e.g., aortic) pulse wave analysis mainly focuses on cardiac conditions, cerebral pulse wave analysis would be concerned with local as well as downstream conditions in the brain. In particular, small arterioles and capillaries, which are too small to be resolved directly by MRI, might affect the pulse waveforms in the main cerebral arteries via reflection or impedance mismatch effects. This way, measuring cerebral pulse waves via MRI hypersampling might open up new avenues for clinical brain imaging.

### C. EEG source localization

The EEG signal often consists of waveforms with a periodic or pseudo-periodic rhythm [2]. Some of the more persistent rhythms, such as alpha, beta, and delta waves, might allow for a similar analysis as the pulse waveform analysis described here. Hypersampling of the MRI signal would be performed as above but with a simultaneously acquired EEG signal [34] as the reference. A possible goal would be to localize the sources of these waveforms in the brain. A necessary condition for the success of such an investigation would be that the MRI signal actually reflects these relatively fast oscillations. There has been recent evidence that this might indeed be the case [35]. Such an investigation is an interesting topic for future research.

### D. Hybrid systems

Hypersampling by APP might be of use for hybrid imaging systems that have a fast and a slow sampling component. For example, the combination of MRI with optical imaging methods such as multiphoton microscopy [36] would provide a slowly sampled MRI signal and a faster sampled optical signal. The optical signal can then be used as the reference signal to detect the fast component in the MRI signal via APP.

### E. Pseudo-periodic and monocomponent signals

On a side note, this manuscript also connects the concepts of monocomponent [37] and pseudo-periodic [38] signals. Based on the here presented application, it is conjectured that monocomponent signals, if properly defined, can become equivalent to pseudo-periodic signals. To what degree monocomponent and pseudo-periodic signals can be seen as equivalent will be left as an exercise for the mathematical sciences [39], though. The wonderful paper of Cohen [37] may serve as a starting point. Finally, it would be desirable if the monocomponent signal requirement could be relaxed to certain multicomponent signals, while retaining the concept of instantaneous frequency [40]. For example, the pulse signal often seems to consist of two components, of which one has been filtered out here by low-pass filtering.

### VI. Conclusion

An algorithm to upsample an insufficiently sampled pseudo-periodic signal with the help of a reference signal,

"hypersampling" has been presented. Hypersampling is based on a projection of the signal time series to the analytic phase of the reference signal. It generalizes the widely used retrospective gating approach. Hypersampling has been validated in numerical simulations of pseudo-periodic signals, and it has been discussed that in the case of nonlinear phase evolution, hypersampling can outperform retrospective gating. Finally, hypersampling has been applied to dynamic MRI data of the human brain, showing a detailed MRI-measured pulse waveform traversing the brain. Possible further applications include source localization of EEG patterns and hybrid imaging systems.

### VII. Acknowledgments and disclosure

The author would like to thank M. Hanke, A. Brechmann, and J. Stadler for help in using their database (M. Hanke et al., Scientific Data 1, 140003, 2014). The author is partially funded by NIH grant # 5 R21 EY027568 02. The study sponsor was not involved in the study design, in the collection, analysis and interpretation of data; in the writing of the manuscript; or in the decision to submit the manuscript for publication. This research uses an algorithm described in a patent application by the Center for Technology Licensing at Cornell University (CTL), "Upsampling of signals by analytic phase projection."

### VIII. Appendix

A script that reproduces Fig. 2. This script can be executed using MATLAB R2017a (The MathWorks, Inc., Natick, MA) with the Signal Processing Toolbox by copying it into the Matlab editor.

```
% Numerical simulation of Analytic Phase Projection.
% This script accompanies the article "Hypersampling of pseudo-
% periodic signals by analytic phase projection" by H.U. Voss,
% published in Computers in Biology and Medicine (2018)

% Definitions

% Signal x(t)
dT=2;
N=200;
period_x=exp(1)/3; % Pseudo-period of irregular cyclic data
timeaxis_x=transpose((0:N-1)*dT);
f_x=1/period_x; % Frequency of wave
T=N*dT; % Common end time for x, xU, y
f0=f_x; % Chirp signal frequencies
f1=2*f0;
x=chirp(timeaxis_x,f0,T,f1,'linear');
x=sin(0.7*pi*x); % Define double peaks
rng(0);
x=x+0.1*std(x)*randn(N,1); % Add noise

% Upsampled time axis in order to see waveforms.
% Must not be used for analysis
dt=0.01; % Sampling time, same as for y(t)
```

```
Ufactor=round(dT/dt); % Upsampling factor
dTU=dT/Ufactor; % 2 ms
NU=N*Ufactor;
timeaxis_xU=transpose((0:NU-1)*dTU);
xU=chirp(timeaxis_xU,f0,T,f1,'linear'); % Upsampled signal
xU=sin(0.7*pi*xU); % Double peaks by nonlin. transform.

% Reference data y(t).
% The same frequency as in the signal has to be used
n=N*dT/dt;
timeaxis_y=transpose((0:n-1)*dt);
y=chirp(timeaxis_y,f0,T,f1,'linear');
y=sin(0.9*pi*y); % Double peaks. Eq. can differ from x(t)

% Computational part

y=(y-mean(y))/std(y); % Normalization

% Step 1: Obtain monocomponent signal
filHz=3;
ts=timeseries((0:n-1),(0:n-1)*dt); % timeseries(data,time)
ts.Data=y;
tmp=idealfilter(ts,[filHz,1e4],'notch');
y_f=tmp.Data;

% Step 2: Computation of analytic signal
xh=imag(hilbert(y_f));
phi_f=atan2(xh,y_f);

% Upsampling factor equals #cycles
numberofcycles=sum(abs(phi_f(2:end)-phi_f(1:end-1))>1.8*pi);
RR=N*dT/numberofcycles;
effectivedT=RR/N;
disp(['Upsampling factor = ' num2str(dT/effectivedT)])

% Step 3: Analytic phase association
inter_phi_f=interp1(timeaxis_y,phi_f,timeaxis_x)';

% Step 4:Creating phase axis (sorting phases)
[phi_f_projected, index]=sort(inter_phi_f);

% Step 5: Analytic phase projection
xAPP=x(index);

% Step 6: Temporal smoothing
xAPP0=xAPP;
smoo_par=1/60;
ts2=timeseries((0:N-1),(0:N-1)*dT); % timeseries(data,time)
ts2.Data=xAPP;
meanvalue=mean(ts2.Data);
tmp=idealfilter(ts2,[smoo_par,1e4],'notch');
xAPP=tmp.Data+meanvalue;

% Figure 2 in manuscript
figure('position',[100.,200.,2*300,2*400], 'PaperOrientation','portrait');

phi_axis=(timeaxis_x/timeaxis_x(end)-.5)*2*pi;
```

```
xAPP0_plot=xAPP0;
xAPP_plot=xAPP;
% Correct for arbitrary phase shift
xAPP0_plot=circshift(xAPP0_plot,20);
xAPP_plot=circshift(xAPP_plot,20);

subplot(4,2,[1,2])
plot(timeaxis_x,x,'k.-','markersize',10)
axis tight
title('Time series x(t)')
xlabel('time t [s]')
ylabel('x(t) [a.u.]')

subplot(4,2,3)
range1=35:35+90-1;
plot(timeaxis_xU(range1),xU(range1),'k-','linewidth',1.5);
axis tight
title('Underlying signal, early')
xlabel('time t [s]')
ylabel('x(t) [a.u.]')
hold on

subplot(4,2,4)
range2=(length(xU)-90+1):length(xU);
plot(timeaxis_xU(range2),xU(range2),'k-','linewidth',1.5);
axis tight
title('Underlying signal, late')
xlabel('time t [s]')
ylabel('x(t) [a.u.]')
hold off

subplot(4,2,5)
range1=1:Ufactor;
plot(timeaxis_y(range1),y(range1),'k-');
hold on
plot(timeaxis_y(range1),y_f(range1),'b-');
plot(timeaxis_y(range1),phi_f(range1),'r-');
range3=1:dT;
plot(timeaxis_x(range3),inter_phi_f(range3),'k*');
axis tight
title('Phase projection, early')
xlabel('time t [s]')
ylabel('[a.u.]')
hold on

subplot(4,2,6)
range2=(length(y)-Ufactor+1):length(y);
plot(timeaxis_y(range2),y(range2),'k-');
hold on
plot(timeaxis_y(range2),y_f(range2),'b-');
plot(timeaxis_y(range2),phi_f(range2),'r-');
range4=N:N;
plot(timeaxis_x(range4),inter_phi_f(range4),'k*');
axis tight
axis tight
legend('Ref. sig. y(t)','Filtered y(t)', ...
    'Phase \Phi(t)', 'Time series x(t)')
```

```
title('Phase projection, late')
xlabel('time t [s]')
ylabel('[a.u.]')
hold off

subplot(4,2,7)
plot(phi_axis,xAPP0_plot,'k.')
axis tight
xlabel('cycle phase \Phi [rad]')
ylabel('[a.u.]')
title('Hypersampled signal x_{APP}(\Phi)');
hold off

subplot(4,2,8)
plot(phi_axis,xAPP_plot,'k.')
axis tight
xlabel('cycle phase \Phi [rad]')
ylabel('[a.u.]')
title('Cycle estimate x"_{APP}(\Phi)');
hold off
```